\documentclass[twoside,a4paper]{article}
\usepackage{fleqn,epsf,graphicx}
\usepackage{authblk}

\begin{document}

\title{Beyond quantum mechanics? Hunting the 'impossible' atoms  (Pauli Exclusion Principle violation and spontaneous collapse of the wave function at test)}

\author{K. Piscicchia$^{a}$$^{b}$

C. Curceanu$^a$$^{b}$, S. Bartalucci$^a$, A. Bassi$^c$, S. Bertolucci$^d$, C. Berucci$^a$$^e$, A. M. Bragadireanu$^a$$^f$, M. Cargnelli$^e$, A. Clozza$^a$, L. De Paolis$^a$, S. Di Matteo$^g$, S. Donadi$^c$, A. d'Uffizi$^a$, J-P. Egger$^h$, C. Guaraldo$^a$, M. Iliescu$^a$, T. Ishiwatari$^e$, M. Laubenstein$^i$, J. Marton$^e$, E. Milotti$^c$, D. Pietreanu$^a$$^f$ T. Ponta$^g$, E. Sbardella$^a$, A. Scordo$^a$, H. Shi$^a$, D.L. Sirghi$^a$$^f$, F.~Sirghi$^a$$^f$, L. Sperandio$^a$, O. Vazquez Doce$^l$, J. Zmeskal$^e$}

\affil{
       $^{a}$INFN Laboratori Nazionali di Frascati,  Frascati (Roma), Italy \\
       $^{b}$Museo Storico della Fisica e Centro Studi e Ricerche "Enrico Fermi", Roma, Italy \\
       $^{c}$Dipartimento di Fisica, Universit\`{a} di Trieste and INFN Sezione di Trieste, Trieste, Italy\\
       $^{d}$CERN, CH-1211,  Geneva 23, Switzerland\\
       $^{e}$The Stefan Meyer Institute for Subatomic Physics, Vienna, Austria\\
       $^{f}$"Horia Hulubei" National Institute of Physics and Nuclear Engineering,  Bucharest - Magurele, Romania\\
       $^{g}$Institut de Physique UMR CNRS-UR1 6251, Universit\'e de Rennes1, Rennes, France\\
       $^{h}$Institut de Physique, Universit\'e de Neuch\^atel, Neuch\^atel, Switzerland\\
       $^{i}$INFN, Laboratori Nazionali del Gran Sasso, Assergi (AQ), Italy\\
       $^{l}$Excellence Cluster Universe, Technische Universit\"at, M\"unchen, Garching,Germany}

\maketitle                   


\begin{abstract}

The development of mathematically complete and consistent models solving the so-called "measurement problem", strongly renewed the interest of the scientific community for the foundations of quantum mechanics, among these the Dynamical Reduction Models posses the unique characteristic to be experimentally testable.
In the first part of the paper an upper limit on the reduction rate parameter of such models will be obtained, based on the analysis of the X-ray spectrum emitted by an isolated slab of germanium and measured by the IGEX experiment.

The second part of the paper is devoted to present the results of the VIP (Violation
of the Pauli exclusion principle) experiment and to describe its recent upgrade.
The VIP experiment established a limit on the probability that the Pauli Exclusion Principle (PEP) is violated
by electrons, 
using the very clean method of searching
for PEP forbidden atomic transitions in copper.

\end{abstract}

\section{Upper limit on the wave function collapse mean rate parameter $\lambda$}

The first consistent and satisfying  Dynamical Reduction Model, known as Quantum Mechanics with Spontaneous Localization (QMSL) \cite{ghi}, considers particles undergoing spontaneous localizations around definite positions, following a Possion distribution characterized by a  mean frequency   $\lambda =10^{-16}$ s$^{-1}$. This brought to the development of the CSL theory \cite{pear} based on the introduction of new, non linear and stochastic terms, in the Shr\"odinger equation besides to the standard Hamiltonian. Such terms induce, for the state vector, a diffusion process, which is responsible for the wave packet reduction.  
As demonstrated by Q. Fu \cite{fu} the particle interaction with the stochastic field also causes an enhancement of the energy expectation value. This implies, for a charged particle, the emission of electromagnetic radiation (known as spontaneous radiation)  not present in the standard quantum mechanics.
The radiation spectrum spontaneously emitted by a free electron was calculated by Fu \cite{fu} in the framework of the non-relativistic CSL model, and it is given by:
$\frac{d\Gamma (E)}{dE} = \frac{e^2 \lambda}{4\pi^2 a^2 m^2 E}$,
where $m$ represents the electron mass, $E$ is the energy of the emitted photon, $\lambda$ and $a$ are respectively the reduction rate parameter and the correlation length of the reduction model which is assumed to be $a=10^{-7}$ m. If the stochastic field is assumed to be coupled to the particle mass density (mass proportional CSL model) (see for example \cite{bassi}) then the previous expression for the emission rate is to be multiplied by the factor $(m_e/m_N)^2$, with $m_N$ the nucleon mass. Using the measured radiation appearing in an isolated slab of Germanium \cite{miley} corresponding to an energy of 11 KeV, Fu obtained the limit $\lambda \leq 0.55 \cdot 10^{-16} s^{-1}$.
In Ref. \cite{Adler} the author argues that, in evaluating his numerical result, Fu uses for the electron charge the value $e^2=17137.04$, whereas the standard adopted Feynman rules require the identification $e^2/(4\pi)=17137.04$. We took into account this correction when evaluating the new limit.  

\begin{figure}[htb]
\centerline{%
\includegraphics*[height=9cm,width=9cm]{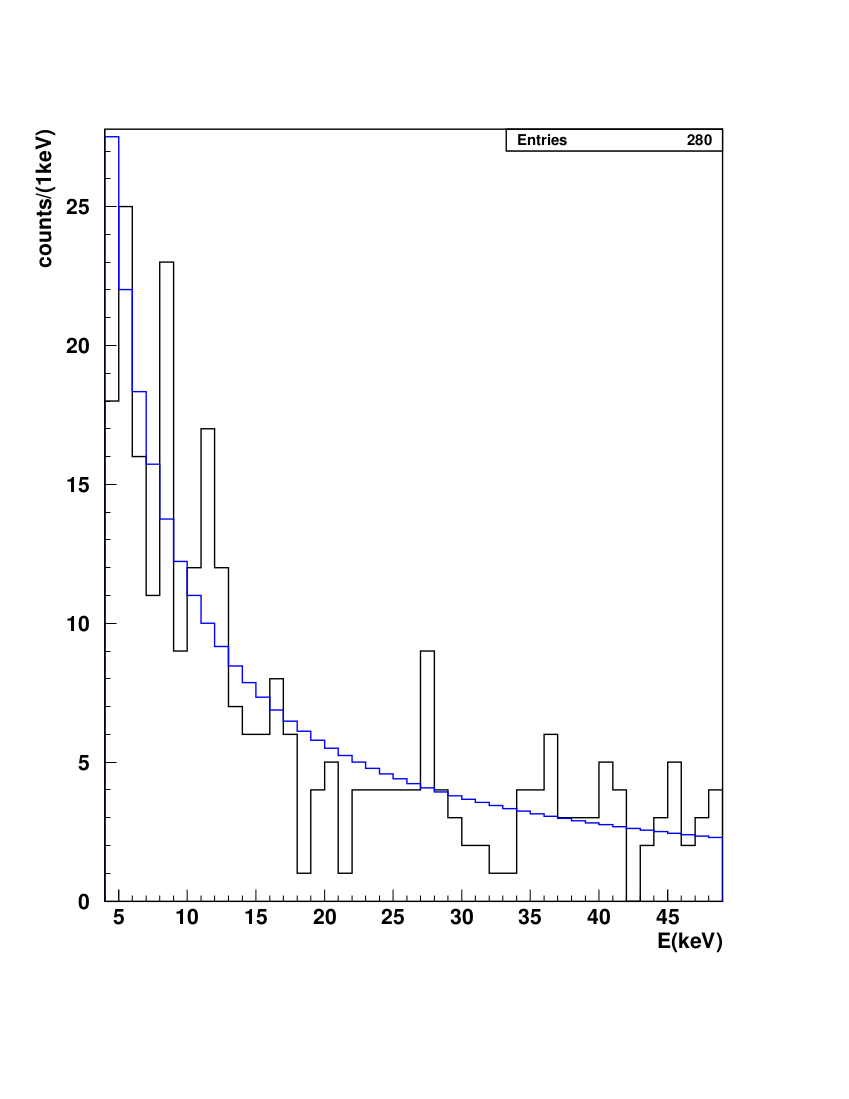}}
\caption{\em
Fit of the X ray emission spectrum measured by the IGEX experiment \cite{igex1,igex2}, performed assuming the predicted energy dependence $\frac{d\Gamma (E)}{dE} = \frac{\alpha(\lambda)}{E}$.}
\label{fit}
\end{figure}

In order to reduce possible biases introduced on the $\lambda$ value by the punctual evaluation of the rate at one single energy bin, the X-ray emission spectrum measured by the IGEX experiment \cite{igex1,igex2} was fitted in the range $\Delta E =$ $4.5\div 48.5$ KeV $\ll m$, compatible with the non-relativistic assumption (for electrons) used in the calculation of the predicted rate.
A Bayesian model was adopted to calculate the $\chi^2$ variable minimized to fit the X ray spectrum, assuming the predicted energy dependence $\frac{d\Gamma (E)}{dE} = \frac{\alpha(\lambda)}{E}$. The result of the performed fit is shown in Figure \ref{fit}.
The minimization gives for the free parameter of the fit the value $\alpha(\lambda) = 110 \pm 7$, corresponding to a reduced chi-square $\chi^2/n.d.f = 1.1$, from which the following upper limits can be set for the $\lambda$ parameter: $\lambda \leq 1.4 \cdot 10^{-17} s^{-1}$ (non mass proportional) and $\lambda \leq 4.7 \cdot 10^{-11} s^{-1}$ (mass proportional).
The obtained limits improve the precedent Fu's limit by a factor 4. Our results are to be compared with the values originally assumed in the models \cite{ghi}:
$\lambda_{QMSL} = 10^{-16} s^{-1},   \lambda_{CSL} = 2.2 \cdot 10^{-17} s^{-1}$
and with the values proposed, more recently, by S. Adler \cite{Adler}. There is still plenty of space to investigate the collapse theory and its consequences.

\section{The VIP experiment and VIP upgrade}

The PEP is a consequence of the
spin-statistics connection \cite{pauli}, and, as such,
 it is intimately connected to the basic axioms of quantum field theory \cite{LZ}.
Given its basic standing in quantum theory, it is  appropriate
to carry out precise tests of the PEP validity and, indeed,
mainly in the last 20 years, several experiments have been performed
to search for possible small 
violations 
\cite{arnold}.

The VIP experiment is dedicated to the measurement of the PEP violation probability for electrons. VIP uses a method developed by  
Ramberg and Snow \cite{ramberg} (in agreement with  the Messiah-Greenberg superselection rule \cite{messiah})
consisting
in the introduction of ``fresh'' electrons into a copper strip, by
circulating a current, and in the search for the X-rays resulting from
the PEP forbidden $2P \to 1S$ ($K_{\alpha}$) transitions
that occur if one
of these electrons is captured by a copper atom and cascades
to a 1S state already filled by two electrons.
The energy of this non-Paulian transition would differ from the normal $K_{\alpha}$ transition energy by
about 300 eV (7.729 keV instead of 8.040 keV) \cite{sperandio}.
The background is evaluated alternating periods without current in the copper strip.

\begin{figure}[htb]
\centerline{%
\includegraphics*[height=9cm,width=15cm]{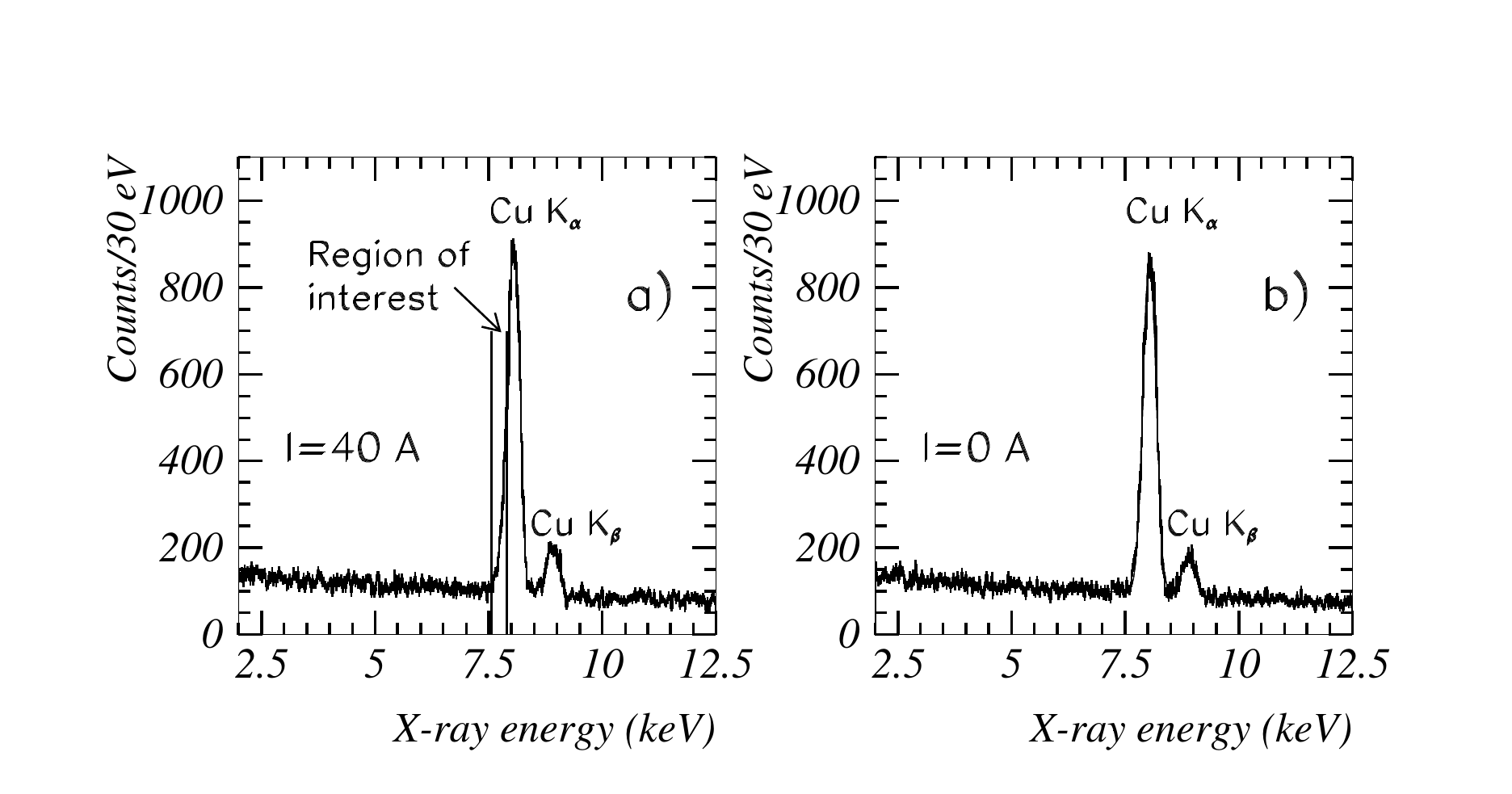}}
\caption{\em
Energy spectra with the VIP setup at LNF-INFN: (a) with current (I = 40 A); (b) without current (I = 0 A).}
\label{spectru}
\end{figure}

The VIP Collaboration set up a much improved version of the Ramberg and
Snow experiment, with a higher sensitivity apparatus \cite{vip}. The detector is an array of 16 Charge-Coupled Devices (CCDs) \cite{culhane}, 
characterized by excellent background rejection capability, based on pattern 
recognition and good energy resolution (320 eV FWHM at 8 keV in the
present measurement). The background was reduced by a careful choice of the
materials and sheltering the apparatus in 
the LNGS underground laboratory of the Italian Institute for Nuclear Physics
(INFN). 
The VIP setup was taking data in a test period at LNF-INFN in 2005,
and the resulting energy calibrated X-ray spectra, for the data taking with and without current, are shown in Figure \ref{spectru}. These spectra include data from 14 CCD's out of 16, because of noise
problems in the remaining 2. Both spectra, apart from the continuous
background component, display clear Cu $K_\alpha$ and $K_\beta$ lines due to X-ray
fuorescence caused by the cosmic ray background and natural radioactivity.
No other lines are present and this reflects the careful choice of the materials
used in the setup, as for example the high purity copper and aluminium, the
last one with $K$-complex transition energies below 2 keV.
The setup was then installed in the Gran Sasso underground
laboratory of INFN where it took data from 2006 until 2010.

To extract the experimental limit on the probability that PEP is violated for electrons, ${\beta^{2}}/{2}$,
from our data, we used the same arguments of Ramberg and Snow: see references
\cite{ramberg} and  \cite{vipart} for details of the analysis. The analysis of the LNGS data gives \cite{tesi,curceanu1}:
$\frac{\beta^{2}}{2} < 4.7 \cdot 10^{-29}$,
which represents an improvement with respect to the previous Ramberg and Snow limit ($\frac{\beta^{2}}{2} <  1.7\times10^{-26}$) of a factor $\sim$ 300.

\begin{figure}
\centering
\includegraphics[height=3.3in]{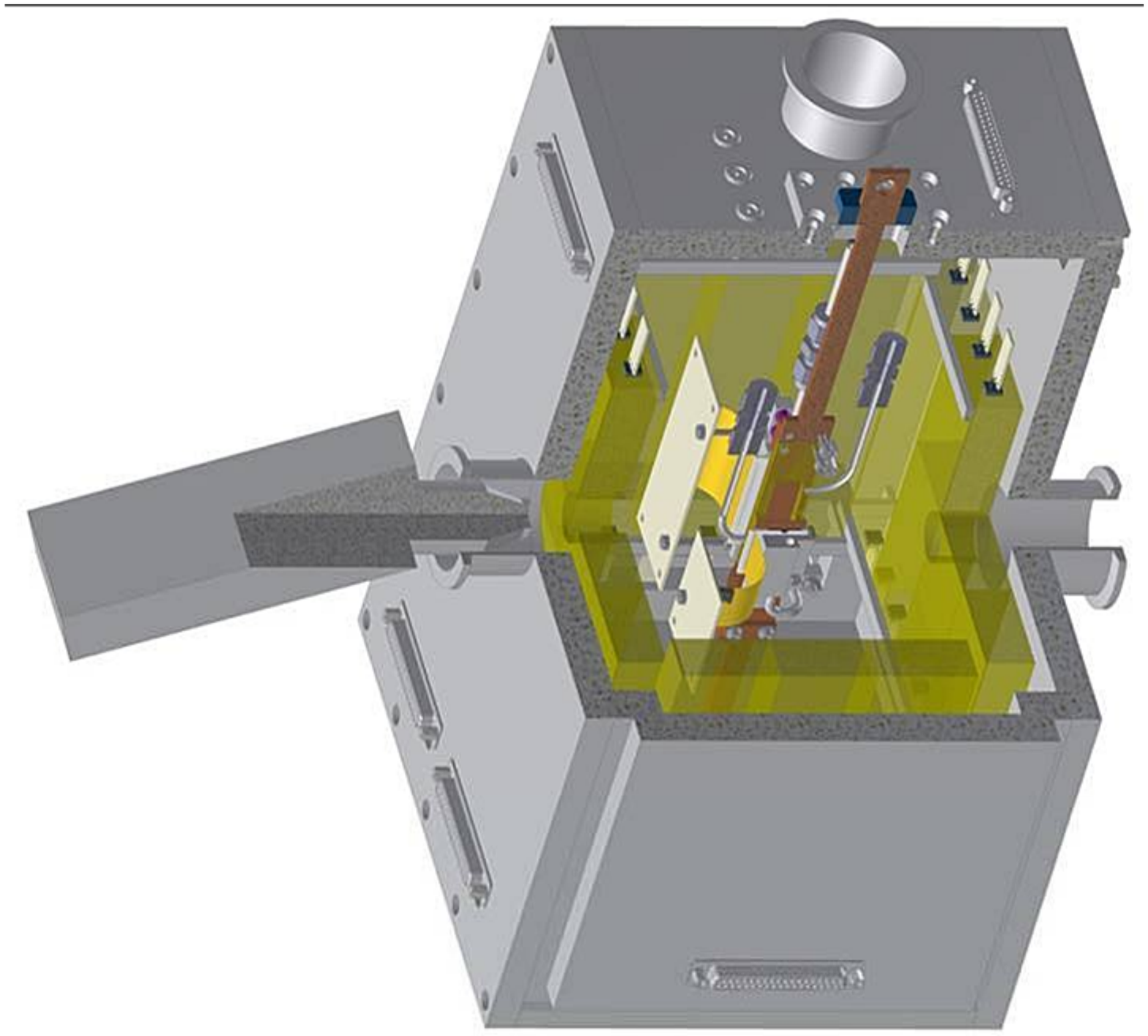}
\caption{\em The schematic implementation of the upgrade of the VIP experiment using
SDD detectors and an external veto-system made of scintillators.}
\label{setup}
\end{figure}

An improved version of the VIP setup was already tested at the LNF and will be installed in the LNGS in next months. Thanks to the substitution of CCDs with the triggerable Silicon 
Drift Detectors (SSD), characterized by a fast readout time ($\simeq 1\mu$s) and large
 collection area (100 mm$^2$), a more compact system was realized, which is shown in Figure \ref{setup}. Moreover to further reduce the background an external veto-system, which would eliminate a large part of the 
background produced by  charged particles coming from the outside the setup, was employed. We expect to gain other 2 orders of magnitude in the limit on the probability of PEP violation, bringing it in the 10$^{-31}$ range.

\end{document}